\documentclass[pre,twocolumn,showpacs,preprintnumbers,floatfix,superscriptaddress,showpacs]{revtex4}
\usepackage{graphicx,citesort,amsmath,amssymb}
\usepackage[dvips]{color}

\renewcommand{\vec}[1]{\pmb{#1}}
\newcommand{\avg}[1]{\left\langle #1\right\rangle}

\newcommand{\abs}[1]{\left\lvert #1\right\rvert}
\newcommand{\mint}[4]{\int_{#2}^{#3}\!\!#1\,#4}

\newcommand{\Ord}[1]{{\cal O}(#1)}

\newcommand{\pe}{\perp}
\newcommand{\pa}{\parallel}

\newcommand{\rpa}{r_\parallel}

\newcommand{\fe}{\mathfrak f \,}

\newcommand{\fel}{\vec{f}_{\text{el}}\,}
\newcommand{\gex}{\mathfrak g \,}

\newcommand{\gfl}{\vec{ g}_{\text{fr}}}
\newcommand{\gel}{\vec{ g}_{\text{el}}}

\newcommand{\Hext}{{\cal H}_{\text{ext}}}
\newcommand{\Hb}{{\cal H}_{\text{WLC}}}

\newcommand{\Htot}{{\cal H}}

\newcommand{\lp}{\ell_p}
\newcommand{\lpe}{\ell_\perp}
\newcommand{\lpa}{\ell_\|}

\newcommand{\tf}{t_\fe}

\newcommand{\pull}{\text{\emph{Pulling}}}

\newcommand{\lelo}{\text{\emph{Release}}}

\preprint{LMU-ASC 33/06}

\begin{document}
\bibliographystyle{apsrev}

\title{Tension dynamics in semiflexible polymers. \\Part~I:
  Coarse-grained equations of motion}

\author{Oskar Hallatschek} \email{ohallats@fas.harvard.edu}
 \affiliation{Lyman Laboratory of Physics, %
   Harvard University, Cambridge, Massachusetts 02138, USA}
 
 \author{Erwin Frey} \affiliation{Arnold Sommerfeld Center for Theoretical Physics
   and Center for NanoScience, %
   LMU M\"unchen, Theresienstr.~37, 800333 M\"unchen,
   Germany}

\author{Klaus Kroy} 
\affiliation{Institut f\"ur Theoretische Physik,%
  Universit\"at Leipzig, Augustusplatz 10/11, 04109 Leipzig, Germany}

\date{\today}

\begin{abstract}
  Based on the wormlike chain model, a coarse-grained description of the nonlinear
  dynamics of a weakly bending semiflexible polymer is developed. By means of a
  multiple scale perturbation analysis, a length-scale separation inherent to the
  weakly-bending limit is exploited to reveal the deterministic nature of the
  spatio-temporal relaxation of the backbone tension and to deduce the corresponding
  coarse-grained equation of motion.  From this partial integro-differential
  equation, some detailed analytical predictions for the non-linear response of a
  weakly bending polymer are derived in an accompanying paper
  (Part~II~\cite{hallatschek-part2:2006}).
\end{abstract}


\maketitle

\section{Introduction}
\label{sec:intro}

The laws of Brownian motion have played the role of a mediator between the apparently
smooth deterministic dynamics on a macroscopic scale and the microscopic molecular
chaos since their discovery a century ago~\cite{FreyK05}. They are pivotal to our
understanding of a broad class of animate and inanimate soft condensed matter systems
that owe their characteristic softness to low-dimensional and strongly fluctuating
meso-scale structures such as polymeric networks and membranes~\cite{GardelSMMMW04}.
Conversely, these systems are well suited to study how complex deterministic dynamics
on a macro- or meso-scale emerges from the underlying stochastic differential
equations~\cite{baschnagel:00}.

\begin{figure}
  \centerline{\includegraphics[width=\columnwidth] {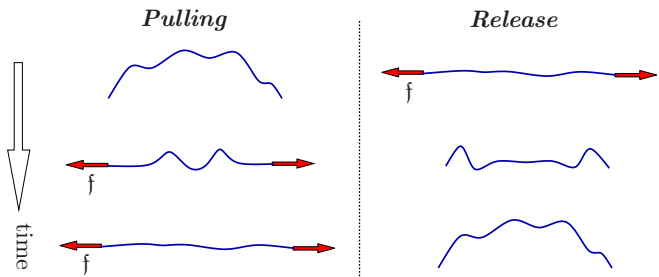}}
  \caption{Two basic examples of dynamic force-extension
    experiments. {\pull}: a weakly bending polymer in equilibrium is
    suddenly pulled longitudinally by two external forces
    $\fe$. {\lelo}: a pre-stretched
    polymer is suddenly released.}
  \label{fig:pull-release}
\end{figure}

Take, for example, a stiff polymer like actin that is suddenly stretched by strong
forces applied at its ends. Or, conversely, consider a polymer that is held in a
virtually straight conformation and suppose that the forces at its ends are suddenly
released.  These two paradigmatic experimental setups, which we call \pull\/ and
\lelo, are illustrated in Fig.~\ref{fig:pull-release}.  How will the end-to-end
distance of the polymer relax to its new equilibrium value?  Given the manifestly
stochastic underlying dynamics, which for \lelo\/ is exclusively driven by thermal
forces, it is not immediately obvious that the initial contraction or stretching
dynamics should obey a deterministic law, as tacitly assumed by several heuristic
derivations~\cite{seifert-wintz-nelson:96,ajdari-juelicher-Maggs:97,morse:98II,everaers-Maggs:99,brochard-buguin-de_gennes:99}.
Indeed, these studies, which predicted a variety of interesting new dynamic scaling
regimes, arrived at partially contradicting
results~\cite{seifert-wintz-nelson:96,brochard-buguin-de_gennes:99}.  Despite
considerable experimental, theoretical and numerical work, both relaxation laws (for
\pull\/ and \lelo) remained controversial for some time.

It therefore appears worthwhile to undertake a detailed mathematical derivation of
the meso-scale dynamic equations that govern the nonlinear dynamics of semiflexible
polymers ``from first principles''~\cite{baschnagel:00}; i.e. from the underlying
stochastic differential equations of motion. In a recent Letter, we have outlined
such a systematic approach that resolves the aforementioned theoretical problems,
together with some of its consequences~\cite{Hallatschek:F:K::94:p077804:2005}. The
present contribution offers a more comprehensive discussion. In Part~I, an effective
coarse-grained meso-scale description of the dynamics of a semiflexible polymer is
developed by means of a multiple scale theory from the stochastic differential
equations of motion. Our detailed analysis also reveals the limits of validity of the
deterministic mesoscopic description and shows how to deal with subtle end effects
that may in some cases mask the non-trivial predictions for certain observables.
Building on this general framework, the theory is elaborated for the specific
problems of {\pull} and {\lelo} in Part~II~\cite{hallatschek-part2:2006}, which provides a template for the future
analysis of a variety of related problems with somewhat different boundary/initial
conditions~\cite{obermayer-hallatschek-frey-kroy:tbp,obermayer-hallatschek2:tbp,obermayer-kroy-frey-hallatschek3:tbp}.
Thereby, we corroborate the importance of a regime of homogeneous tension relaxation,
which generally occurs in {\lelo}-experiments, and establish the complete crossover
scenario between the various intermediate scaling regimes. Additionally, in Part~II
some consequences for common observables are worked out in detail to facilitate
experimental verification of the theory.

Before entering a detailed quantitative analysis, it seems useful to summarize the
main ideas on a qualitative level in order to make the remainder more easily
accessible. A characteristic property of semiflexible polymers and many other
fluctuating meso-scale structures in soft condensed matter, is their reduced
dimensionality or slender shape.  It entails the presence of thermally excited
transverse fluctuations of an essentially inextensible backbone. Returning to the
above example, an actin filament is much more susceptible to bending undulations than
to stretching or compressing its backbone. An analogous statement holds holds for
other biopolymers or two-dimensional locally flat objects like membranes and
surfaces.  This suggests to idealize these structures as \emph{undulating
  inextensible manifolds}~\cite{2002nelson}. To be specific, we focus for the
following on the case of a single semiflexible polymer in solution, which seems to be
the simplest paradigmatic example of the more general soft-matter meso-structures
alluded to above. Its equilibrium mechanical properties and conformational statistics
have by now been thoroughly studied theoretically and
experimentally~\cite{frey-kroy-wilhelm-sackmann:97,yamakawa97}.  Both are well
understood in terms of the self-affine roughness of the equilibrium contour
fluctuations within the so-called \emph{wormlike chain model}, which idealizes the
polymer as an inextensible space curve with an energetic cost for
bending~\cite{saito-takahashi-yunoki:67}. Here, we are primarily interested in the
transient non-equilibrium stretching and contraction dynamics of such a wormlike
chain, i.e., in the much less studied problem of how a semiflexible polymer relaxes
to equilibrium after a sudden drastic change in its boundary conditions.  This
question is of considerable fundamental and practical interest alike, e.g.\ for
single-molecule
manipulations~\cite{perkins-smith-chu:97,quake-babcock-chu:97,CollinRJSTB05} and for
understanding and controlling the dynamic response of polymer solutions and networks
such as those constituting the cytoskeleton of biological
cells~\cite{GaborForgacs06012004,JanmeyW04}.

Because of the inextensible backbone, the stretching or contraction dynamics of a
wormlike chain is entirely due to a spatio-temporal re-organization of the contour
length stored in the transverse thermal wrinkles. It is governed by the dynamic
\emph{backbone tension} $f(s,t)$, which is the force that holds the backbone
together. The dynamics is always assumed to be strongly overdamped by solvent
friction, which can be decomposed into transverse and longitudinal components with
respect to the local tangent, in view of the locally rod-like conformation of the
polymer. At first sight, one might suppose that for small enough undulations one can
resort to a formulation of the dynamics solely in terms of transverse modes for which
only transverse friction matters. Actually, for the equilibrium dynamics in the
absence of external forces, the conclusions based on such an assumption are in accord
with a series of experimental
data~\cite{caspi-etal:98,legoff-hallatschek-frey:02,Shusterman:A:G:K::92:p048303:2004,hohenadl-etal:99}.
At a second thought, considering the constraint of inextensibility, it is far from
obvious how the interplay between transverse and longitudinal friction limits the
relaxation after a sudden application or release of external forces as e.g.\ in
\lelo\/ and \pull.  It is the major objective of the present study to resolve this
puzzle for the case of longitudinal forces, while the somewhat more complex issue of
the nonlinear transverse response will be addressed in a separate
contribution~\cite{obermayer-hallatschek2:tbp}.

As a cornerstone of our derivation, we establish a dynamic scale separation between
the scales where transverse and longitudinal friction reign, namely the transverse
and longitudinal \emph{dynamic correlation lengths} $\lpe(t)$ and $\lpa(t)$,
respectively. We demonstrate that $\lpe(t)\ll \lpa(t)$ holds at any time in the limit
of a \emph{weakly bending rod}.  As a central result, we obtain that the tension
varies to leading order only on the large scale $\lpa(t)$, over which the
short-wavelength transverse undulations that dominate the dissipation on the scale
$\lpe(t)$ are self-averaging. The short-scale transverse fluctuations may thus be
said to provide an effective local backbone compliance for the large-scale
longitudinal dynamics. In this way, transverse and longitudinal dynamics are
effectively decoupled, and the microscopic stochastic differential equations can be
reduced, in a controlled way, to a deterministic (integro-differential) equation for
the coarse-grained meso-scale dynamics.

With some care, most of the predictions that emanate from this reduced description
can qualitatively be obtained from a relatively simple scaling analysis, to which we
will occasionally resort in order to promote the intuitive physical understanding of
our analysis. In fact, not only physical insights but also several interesting formal
predictions were originally obtained on the basis of simple scaling arguments.
However, due to some subtleties and pitfalls, the results available in the literature
remained somewhat contradictory. The following development, in particular Part~I of
this article series, aims at settling the corresponding issues conclusively by
employing a controlled perturbation theory instead of evoking scaling assumptions.

As suggested by the introductory examples of dynamic force-extension experiments,
exemplary realizations of the meso-scale dynamics occur in response to highly
localized forces, which we generally (though not invariably) assume to be applied
abruptly at the boundaries.  They initiate universal self-similar relaxation
processes that spread through the polymer, resulting in characteristic power-law
signatures, so-called intermediate asymptotics~\cite{barenblatt:96}, in various
observables. These will be derived analytically in Part~II~\cite{hallatschek-part2:2006} of this article, which is
devoted to solving the effective deterministic meso-scale equations for the tension,
obtained below.  There, we will also consider the consequences of the tension
dynamics on pertinent observables like the (projected) end-to-end distance and the
novel experimental perspectives brought up by our analysis. For the impatient reader,
our Letter~\cite{Hallatschek:F:K::94:p077804:2005} may serve as a quick guide to our
main arguments and results.

The outline of the present part (Part~I), is as follows. We develop
the systematic coarse grained description of tension dynamics in
stiff, respectively, stretched semiflexible polymers that
emerges from an appropriate small gradient expansion of the dynamical
wormlike chain model (Sec.~\ref{sec:dwmc}).  An ordinary perturbation
expansion leading to linear first order equations of motion
(Sec.~\ref{sec:linear-dynamics}) turns out to be restricted to times
when the tension has already relaxed to its equilibrium value
(Sec.~\ref{sec:intro-tensprop-sfp}). The actual tension dynamics on
shorter times is resolved by a stochastic multiple scale perturbation
theory (Sec.~\ref{sec:multi-scale}), which is based on a dynamical
length scale separation.  As a major result, we obtain a rigorous
deterministic partial integro-differential equation (PIDE) that
describes the long wave-length (-all time-) dynamics of the tension.

\section{Dynamical  wormlike chain model}
\label{sec:dwmc}

At low Reynolds numbers, the dynamics of a polymer is determined by the balance of
elastic forces, friction, and stochastic forces. We will briefly motivate how these
forces are modeled in the usual stochastic description of the Brownian motion of a
semiflexible polymer, leading to Eq.~(\ref{eq:eom-sfp}) below, which is the basis of
our subsequent analysis.

The natural model for the description of the elastic properties of a semiflexible
polymer arises from the idea to regard the polymer as a thin cylinder consisting of a
homogeneous elastic material.  In the slender limit, where the ratio of thickness
over total length approaches zero, the deformation modes of the cylinder become much
stiffer in the axial direction (``phonons'') than in directions transverse to the
cylinder axis (``bending modes'')~\cite{landau-lifshitz-7}.  Consequently, such a
slender rod or ``thread'' subject to external forces merely allows for bending
deformations, while its contour length is to a good approximation locally conserved.

These features are idealized in the wormlike chain model, where the
polymer is at any time $t$ represented as an inextensible space curve
$\vec r(s,t)$ parameterized by the arc length $s$, i.e.\ the tangents
have to obey the constraint
\begin{equation}
  \label{eq:inextensibility}
  \vec r'^2=1 \;.
\end{equation}
Here, we have introduced the shorthand notation $\vec r'\equiv
\partial \vec r(s,t)/\partial s$. The effective free energy $\Hb$ of a
particular conformation is only due to bending (curvature) and is
given by
\begin{equation}
  \label{eq:bending-energy}
  \Hb= \frac\kappa 2\mint{ds}{0}{L} \vec r''^2 \;,
\end{equation}
where $\kappa$ is the bending stiffness. To assure that the integrand is the square
of the local curvature, the inextensibility constraint,
Eq.~(\ref{eq:inextensibility}), has to be enforced as a rigid constraint.

The elastic force (per arclength) $\gel$ derives from Eq.~(\ref{eq:bending-energy})
by a functional derivative,
\begin{equation}
  \label{eq:elastic-force-general}
  \gel(s,t)=-\left. \frac{\delta \Hb}{\delta \vec r }\right|_{\vec r'^2=1} \;.
\end{equation}
As indicated, the contour variations $\delta \vec r$ used to determine the functional
derivative on the right-hand-side have to respect the local inextensibility
constraint, Eq.~(\ref{eq:inextensibility}). The variational calculation, detailed in
App.~\ref{sec:elastic-forces}, yields
\begin{equation}
  \label{eq:beq}
  \vec{r}' \times \left( \kappa\, \vec{r}'''+\fel
      \right)= 0 \;,
\end{equation}
where $\fel(s)\equiv \mint{d\tilde{s}}{0}{s} \gel(\tilde{s})$ is the
spatial integral over the elastic force density defined in
Eq.~(\ref{eq:elastic-force-general}). 

According to the implicit equation for $\fel$, Eq.~(\ref{eq:beq}), the force
$\kappa\, \vec{r}'''+\fel$ has a vanishing component transverse to the local tangent
$\vec{r}'$.  Equivalently, we may require that both vectors are proportional to each
other,
 \begin{equation}
  \label{eq:beq-2}
   \kappa\, \vec{r}''' +\fel =  f\, \vec{r}' \;,
\end{equation}
where the proportionality factor $f(s)$ has dimensions of a force and, in general,
depends on the arc length. A spatial derivative then yields a direct expression for
the elastic force density,
\begin{equation}
  \label{eq:beq-3}
  \gel = -\kappa\, \vec{r}'''' +  (f \vec{r}')' \;.
\end{equation}
The apparent simplification with respect to Eq.~(\ref{eq:beq}) is somewhat deceiving,
since Eq.~(\ref{eq:beq-3}) still contains the unknown function $f(s)$, which has to
be fixed by the local arc-length constraint, Eq.~(\ref{eq:inextensibility}). The new
force field $f(s)$ has however a very direct and intuitive physical interpretation.
It is the \emph{local line tension}~\cite{JMDeutsch05131988}  that ``ties the polymer
together'' and thereby enforces the inextensibility condition.

Much of the following deals with the dynamics of this local line tension, which turns
out to be closely related to the dynamics of the local excess length stored in
undulations, defined in Eq.~(\ref{eq:t-t-corr-fct}) below, to which we will refer to
as \emph{stored length}. Time-dependence is introduced into the description by
requiring the elastic force density $\gel$ to be balanced by the dynamic friction
(per arc length) with the solvent,
\begin{equation}
  \label{eq:gex-local-friction}
  \gfl(s,t)=-\vec \zeta(\vec r,t) \partial_t \vec r(s,t) \qquad \text{(free
  draining)}\;.
\end{equation}
At any arclength $s$ and time $t$, the friction matrix $\vec \zeta$
with elements $\zeta_{ij}$ can be decomposed into its transverse and
longitudinal components with respect to the local tangent $\vec
r'(s,t)$,
\begin{equation}
  \label{eq:friction-tensor}
  \vec \zeta=\left[\zeta_\perp \left(1-\vec r'\otimes \vec r'
    \right)+\zeta_\pa \;\vec r' \otimes \vec r' \right]  \;.
\end{equation}
The constants $\zeta_\perp$ and $\zeta_\pa$ can to a first
approximation be estimated by the friction coefficients (per length)
for transverse respectively longitudinal motion of a \emph{rigid}
slender rod in a solvent of viscosity $\eta$~\cite{doi-edwards:86},
\begin{equation}
  \label{eq:tfriction}
  \zeta_\pe=2 \zeta_\pa \simeq 4\pi\eta \;.
\end{equation}
A more sophisticated analysis would consider logarithmic
corrections~\cite{batchelor:70,cox:70a,cox:70b} to account for the dynamic coupling
of distant chain segments $s$ and $s'$ via long-ranged hydrodynamic
interactions~\cite{degennes:79,doi-edwards:86}.  While such logarithmic factors may
sometimes be crucial in a quantitative comparison with some
experiments~\cite{hohenadl-etal:99,legoff-hallatschek-frey:02}, their (feasible) implementation is not of
primary interest to our present discussion, so that we chose to dismiss them for
greater clarity of the presentation.

On the same level of approximation, the force balance of elastic and frictional
forces can be extended by adding thermal white noise fully characterized in terms of
its mean and variance,
\begin{subequations}
  \begin{eqnarray}
    \label{eq:zero-mean}
    \avg{\xi_i(s,t)}&=&0\;,\\
    \label{eq:noise-corrs}
    \avg{ \xi_i(s,t) \xi_j(s',t')}&=&2 k_B T \zeta_{ij} \delta(s-s')
    \delta(t-t')\;
  \end{eqnarray}
\end{subequations}
with the angular brackets indicating an ensemble average. The strength of the noise
correlations in Eq.~(\ref{eq:noise-corrs}) is dictated by the
Fluctuation--Dissipation theorem~\cite{kubo83bookI,kubo91bookII}, which assures that
the steady state of the stochastic equations of motion correspond to thermal
equilibrium (for complications in simulations of discrete bead-rod chains, see
Ref.~\cite{Fixman78,Hinch94,MorseD04}).

Upon using Eqs.~(\ref{eq:beq-3},~\ref{eq:gex-local-friction}), the balance of elastic,
friction and stochastic forces, $0=\gel+\gfl+\vec \xi$, takes the form of an equation
of motion,
\begin{equation}
    \label{eq:eom-sfp}
     \vec \zeta \partial_t \vec r=- \kappa\, \vec r'''' +  (f \vec
    r')'+\vec \xi\;.
\end{equation}
The partial differential equation Eq.~(\ref{eq:eom-sfp}), the arc length constraint
Eq.~(\ref{eq:inextensibility}) and the Gaussian noise-correlation comprise a
complete stochastic description~\footnote{Since the thermal forces have a physical
  origin, Eq.~(\ref{eq:eom-sfp}) has to be interpreted according to
  Stratonovi$\hat{c}$~\cite{vanKampen:book}.} of the Brownian dynamics in the free
draining limit.

\section{Linearized stochastic dynamics}
\label{sec:linear-dynamics}
The nonlinear stochastic dynamics of a wormlike chain, represented by
Eq.~(\ref{eq:eom-sfp}), in combination with the inextensibility constraint,
Eq.~(\ref{eq:inextensibility}), is hard to analyze in general.  The difficulties are
largely due to the calculation of the line tension $f(s,t)$, which has to enforce the
local inextensibility constraint, Eq.~(\ref{eq:inextensibility}).  There have been
attempts to relax the local constraint and merely enforce local inextensibility on
average~\cite{ha-thirumalai:97a,lee-thirumalai:04,harnau-winkler-reineker:95}.  This
corresponds to replacing the field $f(s,t)$ by a (spatially averaged) mean field.
However those approaches fail to describe semiflexible polymers on local scales,
where tension fluctuations (in time and space) are
important~\cite{seifert-wintz-nelson:96}.

\begin{figure}
  \centerline{\includegraphics[width=\columnwidth]
    {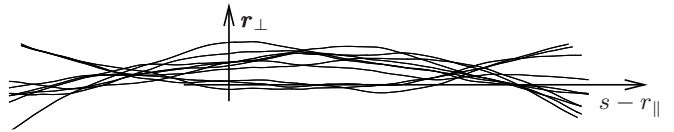}}
  \caption{Typical conformations of a freely fluctuating stiff polymer with
    suppressed rotations. In order to generate these conformations, which serve
    illustrative purposes only, we have represented each conformation as a linear
    superposition of the first $10$ modes that solve the linearized equation of
    motion. For each realization, the mode amplitudes were then drawn randomly
    according to their Gaussian distribution in equilibrium.  }
  \label{fig:flucting-line}
\end{figure}

Our analytical approach is instead based on the weakly bending limit, in which the
polymer conformation is appropriately described by its deviations from a straight
line.  Accordingly, we parameterize the polymer's space curve by 
\[\vec r=(\vec r_\pe,s-r_\pa)^T \qquad\text{(cf.  Fig.~\ref{fig:flucting-line}),}\] where $\vec r_\pe(s)$
and $r_\pa(s)$ are the transverse and longitudinal displacements at arc length $s$.
Such parameterization is only useful if the gradients of the transverse displacements
are small everywhere along the contour,
\begin{equation}
  \label{eq:wbl-sfp}
    \vec r_\pe'^2=\Ord{\epsilon}\ll 1 \;,\qquad s\in (0,L)\;. 
\end{equation}
The weakly bending limit, as defined in Eq.~(\ref{eq:wbl-sfp}),
assumes that there is a small parameter $\epsilon$ that controls the
polymer roughness \emph{uniformly} along the contour. In the case of a
stiff polymer, on which we focus if otherwise stated, the small
parameter is provided by the ratio
\begin{equation}
  \label{eq:wbl-sfp-2}
  \epsilon\equiv \frac L{\lp}=\frac{L k_B T}{\kappa}\ll1 \qquad 
  \text{(stiff polymer)}
\end{equation}
of length $L$ over persistence length $\ell_p$. Alternatively, a weakly bending
conformation may be realized by applying a static external stretching force $\fe$
larger than the internal characteristic force scale $k_B T/\lp$.  In this case, $\vec
r_\pe'^2\sim\sqrt{k_B T/\lp\fe}$~\cite{marko-siggia:95} and we may identify
\begin{equation}
  \label{eq:stretched-p}
  \epsilon\equiv \sqrt{k_B T/\lp\fe}\qquad \text{(stretched polymer).}
\end{equation}

Eq.~(\ref{eq:wbl-sfp}) allows us to expand the dynamical equations of motion in terms
of small gradients. We start with the inextensibility constraint, which ``slaves''
the higher order longitudinal displacements to the transverse ones. After resolving
Eq.~(\ref{eq:inextensibility}) for $r_\pa'$ and expanding the square root, the local
constraint takes the simple form
\begin{equation}
  \label{eq:constraint-wbl-sfp}
  \rpa'=\frac1 2 \vec r_\pe'^2+\Ord{\epsilon^2}=\Ord{\epsilon}\;.
\end{equation}
This entails that the parameter $\epsilon$ is a measure for the reduction of the
longitudinal extension
\begin{equation}
  \label{eq:rpa-def}
  R_\pa\equiv L-r_\pa(L)+r_\pa(0)
\end{equation}
of the polymer due to the presence of thermal undulations: An arclength integral of
Eq.~(\ref{eq:constraint-wbl-sfp}) shows that $R_\pa$ is smaller then the total
contour length $L$ by an amount of the order
\begin{equation}
  \label{eq:contraction}
  L-R_\pa=r_\pa(L)-r_\pa(0)=\Ord{\epsilon L}\;.
\end{equation}
We may think of the length difference in Eq.~(\ref{eq:contraction}) as being
``stored'' in undulations. The distribution of this excess length along the filament
is, according to Eq.~(\ref{eq:constraint-wbl-sfp}), described by the function
\begin{equation}
  \label{eq:t-t-corr-fct}
  \varrho(s,t)\equiv \frac12 \vec{ r}_\pe'^2(s,t)= \Ord{\epsilon}\;,
\end{equation}
which will have central importance in our analysis. It is the fraction of the contour
length that is at arclength $s$ and time $t$ \emph{locally} stored in undulations.

After taking a ``spatial'' (i.e.\ arc-length) derivative of its longitudinal part, we
expand the equation of motion, Eq.~(\ref{eq:eom-sfp}), to order
$\Ord{r_\pe'^2}=\Ord{\epsilon}$  and obtain
\begin{subequations}
  \label{eq:sfeom}
  \begin{eqnarray}
    \label{eq:perp-eq-motion-sfp}
    \partial_t \vec r_\pe&=&-\vec r_\pe''''+ (f \vec
    r_\pe')'+\vec{\gex}_\pe+\boldsymbol \xi_\pe  \\
    \hat \zeta \partial_t r'_\pa&=&(\hat \zeta-1)(\vec r_\pe'
    \partial_t\vec r_\pe)'- r_\pa'''''-f''+ (f r_\pa')''\nonumber\\
    &&-\gex_\pa'-\xi_\pa' 
    \label{eq:parallel-eq-motion-sfp}
    \;.
  \end{eqnarray}
\end{subequations}
Here, we have neglected terms of order $\Ord{\epsilon^{3/2}}$ and made the following
choice of units: time and tension, respectively, are rescaled according to
\begin{eqnarray}
  \label{eq:rescaling-time-tension}
  t&\to&\zeta_\perp t/\kappa  \\
  f&\to& \kappa f\;.  
\end{eqnarray}
This corresponds to setting $\kappa\equiv\zeta_\pe\equiv 1$ and $\hat \zeta\equiv
1/2=\zeta_\pa$. As a consequence all variables represent powers of length, e.g., $t$
and $f$ are a length$^4$ and a length$^{-2}$, respectively. In Eq.~(\ref{eq:sfeom})
we have further allowed for an external~\footnote{Throughout, we reserve Gothics for
  \emph{external} forces or force fields like $\vec \gex$.} force density $\vec \gex=(\vec \gex_\pe,\gex_\pa)$, which is a
length$^{-3}$, that may for instance represent the effect of a solvent flow, an
optical/magnetical tweezer or an electrical field.

As long as one is taking the limit $\epsilon\to 0$ with $t,s,L$ fixed, the polymer is
correctly described by the linearized version of Eq.~(\ref{eq:sfeom}),
\begin{subequations}
  \label{eq:eom-low}
  \begin{align}
    \label{eq:eompe-low}
    \partial_t \vec r_\pe&=-\vec r_\pe''''+ (f \vec
    r_\pe')'+\vec{\gex}_\pe+\vec \xi_\pe  \\
    \label{eq:eompa-low}
    f''&=-\gex_\pa' \;,
  \end{align}
\end{subequations}
From the magnitude of the noise-correlations, given by Eq.~(\ref{eq:noise-corrs}), it
may be inferred that $\vec \xi_\pe$ to leading order obeys
\begin{equation}
  \label{eq:noise-corrs-low}
  \avg{\vec \xi_\pe(s,t) \vec \xi_\pe(s',t')}=2 (\bold I/\lp)\delta(s-s')
  \delta(t-t')\;,
\end{equation} 
where we used $k_B T \zeta_\pe\to\lp^{-1}$ in our choice of units and $\bold
I_{ij}=\delta_{ij}$ is the identity matrix.

From Eq.~(\ref{eq:eompa-low}), it is seen that the tension profile of
a polymer that is forced at the ends is linear; slope and offset are
fixed by the boundary conditions. The (higher order) longitudinal
displacements are slaved to the transverse ones by the arc length
constraint, Eq.~(\ref{eq:constraint-wbl-sfp}). At the present level of
approximation the exact equations of motion have reduced to a linear
equation for the transverse displacements alone.

\subsection{Generalized transverse response}
\label{cha:LT}

In many practical cases, for instance if the polymer is symmetrically
pulled apart by a (possibly time-dependent) force, the tension
$f=f(t)$ is to lowest order spatially homogeneous such that the
equation of motion, Eq.~(\ref{eq:eompe-low}), reduces to
\begin{equation}\label{eq:eom1}
  \partial_t \vec r_\pe = -\vec r_\pe''''+f \vec r_\pe'' + \vec\xi_\pe
  \;.
\end{equation}
Let us anticipate at this point that we will identify an inherent length scale
separation, Eq.~(\ref{eq:lscale-sep}) in Sec.~\ref{sec:multi-scale} below, according
to which the tension can be considered as slowly varying in space, such that
Eq.~(\ref{eq:eom1}) describes the polymer dynamics \emph{locally} (and even globally
at late times). As a consequence, the solution of Eq.~(\ref{eq:eom1}) for a given
spatially homogeneous tension history $f(t)$ becomes an important ingredient of the
nonlinear theory and shall be analyzed in the following.

Formally, the linear Langevin equation Eq.~(\ref{eq:faltung}) is solved by
\begin{equation}
  \label{eq:faltung}
  \vec r_\pe(s,t)=\mint{ds'}{0}{L}\mint{dt'}{-\infty}{\infty}\chi(s,s';t,t')
  \,\vec \xi_\pe(s',t') \;.
\end{equation}
The Green's function $\chi(s,s';t,t')$ satisfies
\begin{eqnarray}
  \label{eq:scb-bc}
  \partial_t\chi(s,s';t,t')&=&-\partial_s^4\chi(s,s';t,t') + f(t)
  \partial_s^2 \chi(s,s';t,t')\nonumber \\ &&+\delta(s-s')
  \delta(t-t')
\end{eqnarray}
and appropriate boundary conditions. It may be interpreted as a causal response
function that describes the spreading and the decay of contour undulations induced by
a transverse force impulse at location $s'$ and (elapsed) time $t'$.
Eq.~(\ref{eq:faltung}) therefore can be said to represent the conformation at time
$t$ in terms of the accumulated response to the transverse noise history $\vec
\xi_\pe(s',t')$ along the contour.

In the general case of a time-dependent tension, it can be quite difficult to
determine the Green's function $\chi$ that obeys the prescribed boundary conditions,
because eigenmodes and eigenvalues of the linear operator
$(f(t)+\partial_s^2)\partial_s^2$ depend on the value of $f(t)$ and thus become
time-dependent.  In terms of Fourier modes, on the other hand, a translationally
invariant Green's function $\overline{\chi}(s-s';t,t')$ can easily be found, below.
As will be detailed in Sec.~\ref{sec:intro-tensprop-sfp}, this function describes the
universal bulk dynamics far away from the ends, while a correction term that
manifestly breaks translational invariance has to be added ``close'' to the ends to
correct for the actual boundary effects.

To formalize this decomposition into bulk and boundaries, the full
response function $\chi$ may be written as a superposition
\begin{equation}
  \label{eq:superpose}
  \chi(s,s';t,t')=\overline{\chi}(s-s';t,t')+\chi_{bc}(s,s';t,t') \;,
\end{equation}
where $\overline{\chi}(s-s';t,t')$ and $\chi_{bc}(s,s';t,t')$ represent a
translationally invariant part and the boundary correction, respectively. The former
is taken to satisfy
\begin{equation}
  \label{eq:trans-inv-eom}
    \partial_t\overline{\chi}(s;t,t')=[-\partial_s^2+f(t)]\partial_s^2\overline{\chi}(s;t,t') +\delta(s) \delta(t-t')
\end{equation}
on an infinite arc length interval. With help of Fourier modes,
\begin{equation}
  \label{eq:FT}
   \overline{\chi}(q;t,t')\equiv \mint{ds}{-\infty}{\infty}
   \overline{\chi}(s;t,t') e^{-i q s} \;,
\end{equation}
Eq.~(\ref{eq:trans-inv-eom}) reads
\begin{equation}
  \label{eq:scb}
  \partial_t\overline{\chi}(q;t,t')+\lambda(q,t)\overline{\chi}(q;t,t') = \delta(t-t') \;,
\end{equation}
where $\lambda(q,t)$ is the dispersion relation~\footnote{The dispersion relation
  Eq.~(\ref{eq:disp-rel}) is easily extended to accommodate a general time dependent
  transverse harmonic confinement potential represented by an additional spring
  constant on the RHS of Eq.~(\ref{eq:disp-rel}).}
\begin{equation}
  \label{eq:disp-rel}
  \lambda(q,t)=q^4+f(t)q^2 \;.
\end{equation}
By the method of integrated factors, the solution to Eq.~(\ref{eq:scb}) is found to
be~\footnote{We note aside, that $\overline{\chi}^2(q;t,0)$ is identical to the
  ``amplification factor'' introduced in Ref.~\cite{hallatschek-frey-kroy:04} to
  describe the growth of the squared mode amplitudes.}
\begin{equation}
  \label{eq:chi-q}
  \overline{\chi}(q;t,t')=\Theta(t-t') \exp\left[-\mint{d\hat t}{t'}{t}\lambda(q,\hat t)\right]\;,
\end{equation}
which may  be checked by direct  substitution. The real  space susceptibility is
given by the inverse Fourier transform of Eq.~(\ref{eq:chi-q}),
\begin{equation}
  \label{eq:chi-real}
  \overline{\chi}( s;t,t')=\mint{\frac{dq}{\pi}}{0}{\infty}\overline{\chi}(q;t,t')\cos(q s) \;,
\end{equation}
where it has been used that $\overline{\chi}(q;t,t')$ is even in $q$.

The part $\chi_{bc}(s,s';t,t')$ of the susceptibility, which is not translationally
invariant, satisfies the homogeneous differential equation
\begin{equation}
  \label{eq:homo-lin-pde}
  \partial_t\chi_{bc}=[-\partial_s^2+f(t)]\partial_s^2 \chi_{bc} \;,
\end{equation}
and has boundary conditions that have to compensate for the generally inappropriate
behavior of $\overline{\chi}(s-s';t,t')$ at the boundaries.  The difficulty of
solving Eq.~(\ref{eq:scb-bc}) has been shifted to $\chi_{bc}(s,s';t,t')$. However,
for the time-dependent quantities to be studied below, this boundary term represents
a relevant contribution only within a characteristic length $\lpe(t)$ (defined below)
close to the boundaries. For times small enough, such that $\lpe(t)\ll L$, one may
use $\chi_{bc}$ derived on a semi-infinite polymer to approximate the situation near
one boundary, say the one at $s=0$.  For simplicity, we will mostly refer to the
model boundary conditions of ``hinged'' (h) or ``clamped'' (c)
ends~\cite{WigginsROG98}, for which the full susceptibility on a semi-infinite
arclength interval $s\in (0,\infty)$ is given by a symmetric and antisymmetric
combination of the bulk susceptibility $\overline{\chi}$,
\begin{equation}
  \label{eq:susceptibility-hinged-clamped}
  \chi_{\frac hc}(s,s';t,t')= \overline{\chi}(s-s';t,t')\mp\overline{\chi}(s+s';t,t')\;.
\end{equation}
Evidently, $\chi_{\frac hc}$ satisfies Eq.~(\ref{eq:scb-bc}) on $s\in (0,\infty)$, as
well as the boundary conditions
\begin{eqnarray}
  \label{eq:bcs-h-c}
  \chi_h(0,s';t,t')=&0&=\partial_s^2\chi_h(0,s';t,t')\qquad
  \text{(hinged)}\nonumber \\
  \partial_s\chi_c(0,s';t,t')=&0&=\partial_s^3\chi_c(0,s';t,t')\qquad
  \text{(clamped)}\;.\nonumber
\end{eqnarray}
A hinged end has vanishing transverse displacement and is torque free
(vanishing second derivative), whereas a (``gliding'') clamped
end has a vanishing slope and is force free (vanishing third
derivative).

\subsection{Local response of the stored excess length }
\label{sec:stored-length}

As a basis for our subsequent systematic analysis of tension dynamics and as an
application of the above results, we wish to determine the longitudinal motion
implied by the linearized transverse stochastic dynamics, Eq.~(\ref{eq:eompe-low}).
To this end, consider a stiff polymer equilibrated under a constant tension $f_<$ at
time zero, on which a \emph{spatially constant} tension $f(t)$ is imposed that varies
\emph{deterministically} for $t>0$.  We ask, how the density of stored excess length
$\varrho(s,t)$, defined in Eq.~(\ref{eq:t-t-corr-fct}), changes in time by
considering the ensemble average of the increase (during the time interval $t$) of
the stored length
\begin{equation}
  \label{eq:change-in-slength}
  \Delta \varrho(s,t)\equiv\varrho(s,t)-\varrho(s,0) \;.
\end{equation}
This is an important observable, since it governs the leading order contribution to
the change \[\Delta R_\pa(t)\equiv R_\pa(t)-R_\pa(0)\] in the projected end-to-end
distance $R_\pa(t)$, defined in Eq.~(\ref{eq:rpa-def}), which is can be directly
measured in dynamic single polymer experiments~\cite{legoff-hallatschek-frey:02}.
Here, the (average) end-to-end axis of the polymer is assumed to be controlled by
external means; e.g.\ by an external force field, flow field, boundary conditions,
etc. To the relevant order, the precise measures taken to orient the polymer
(strictly or on average) do not matter, and we find
\begin{equation}
  \label{eq:end-end-length}
  \avg{\Delta R_\pa} (t)= -\mint{ds}{0}{L}\avg{\Delta\varrho}(s,t)+o(\epsilon)\;.
\end{equation}
Here, the notation with the arguments $s$ and $t$ outside the brackets of
$\avg{\Delta\varrho}$ was introduced to emphasize that, even after averaging, these
dependencies generally persist.

For an explicit calculation of the stored length, we insert Eq.~(\ref{eq:faltung})
into Eq.~(\ref{eq:t-t-corr-fct}) and perform an ensemble average upon employing
Eq.~(\ref{eq:noise-corrs-low})
\begin{eqnarray}
  \avg{\varrho}(s,t)    &= & \mint{dt'}{-\infty}{t}\mint{ds'}{0}{L}\mint{d\tilde
    t'}{-\infty}{t}\mint{d\tilde s'}{0}{L}\partial_s \chi(s,s';t,t')\nonumber\\
  &&\partial_s\chi(s,\tilde s';t,\tilde
  t')\avg{\vec \xi_\pe(s',t') \cdot \vec \xi_\pe(\tilde s',\tilde t')}/2\nonumber \\
  &=& 2 \mint{\frac{dt'}{\lp(t')}}{-\infty}{t}\mint{ds'}{0}{L}\partial_s \chi(s,s';t,t')^2
 \;.  \label{eq:varrho-full} 
\end{eqnarray}
For later convenience, we have allowed for a \emph{time-dependent} persistence length
$\lp(t)$ and an optional prestress $f_<\gg L^2$.  The latter is also technically
advantageous, since it acts as a physical regularization to suppress modes with
wavelength larger than the total length. It enables us to take the total length to
infinity and to discuss the stored length $\avg{\varrho}(s,t)$ on a semi-infinite
arclength interval. For our ultimate goal of calculating $\avg{\Delta \varrho}(s,t)$
an intrinsic regularization renders modes with wave length beyond a characteristic
length scale $\lpe(t)$ irrelevant, so that $f_<$ can eventually be set to zero if
required.

Inserting Eq.~(\ref{eq:susceptibility-hinged-clamped}), valid for hinged/clamped
boundary conditions, into Eq.~(\ref{eq:varrho-full}) (with $L\to\infty$) yields
\begin{eqnarray}
    \avg{\Delta\varrho_{\frac hc}}&=&\mint{\frac{dq}{\pi}}{0}{\infty} 
    \left[ \hat\varrho(q,t)-\hat\varrho(q,0)  \right]    \left[
      1\pm \cos(2 q s)  \right]  \nonumber\\
     \hat\varrho(q,t) &=&2q^2 \mint{d\tilde t}{-\infty}{t}
      \overline{\chi}^2(q;t,\tilde t)/\lp(\tilde t)\;. \label{eq:general-stolength-b}
\end{eqnarray}
The general expression Eq.~(\ref{eq:general-stolength-b}) is now specialized to the
scenario 
\begin{equation}
  \label{eq:force-history}
  \left.
    \begin{aligned}&\quad \text{tension } & \quad\text{persistence length} \\ 
      t<0&:\quad f_< = \text{const.} & \lp= \text{const.}  \qquad \\ t>0&:\quad f(t) & \lp/\theta = \text{const.} \qquad
    \end{aligned}
  \right.
\end{equation}
As preparation for a more general discussion, we have included the possibility of a
sudden change in persistence length by a factor $1/\theta>0$ at $t=0$.

With a constant tension at negative times, the time-integral in
Eq.~(\ref{eq:general-stolength-b}) can be evaluated from $\tilde t=-\infty$ to
$\tilde t=0$, after which we obtain
\begin{eqnarray}
  \label{eq:slength-modes-final}
  \hat \varrho(q,t)\,\lp&=&\frac{\overline{\chi}^2(q;t,0)}{q^2+f_<}+
    2 \theta q^2\mint{d\tilde t}{0}{t} \overline{\chi}^2(q;t,\tilde t)\;.
\end{eqnarray}
Therewith, the full expression for the average change in stored excess length density
becomes
\begin{eqnarray}
  \avg{\Delta\varrho_{\frac hc}}(s,t)&= &\mint{\frac{dq}{\pi\lp}}{0}{\infty}
  \left\{\frac{1}{q^2+f_<}\left(e^{-2q^2[q^2 t+ F(t)]}-1\right)\right. \nonumber\\ 
    & &+\left. 2 \theta
    q^2\mint{d\tilde t}{0}{t}e^{-2q^2\left[q^2(t-\tilde
        t)+F(t)-F(\tilde t)\right]}\right\}\nonumber\\
  & &\left[ 1\pm\cos(2 q s)  \right]\;, \label{eq:delta-rho-mega-expression}    
\end{eqnarray}
where $F(t)$ is the integral of the tension over positive times,
\begin{equation}
  \label{eq:F}
  F(t)=\mint{d\hat t}{0}{t} f(\hat t) \;.
\end{equation}
One observes that the integral in Eq.~(\ref{eq:delta-rho-mega-expression}) is well
defined, even for $f_<=0$, because the integrand vanishes for $q\to 0, \infty$. It is
dominated by wave numbers for which the exponents become of order one, i.e., the
dominant modes have wave numbers $q$ for which the characteristic relaxation time
\begin{equation}
  \label{eq:relaxation-times}
  \tau_q=(q^4 +q^2 F/t)^{-1}
\end{equation}
is of the order of $t$.  This suggests to define a characteristic length $\lpe(t)$ as
the wave length for which the relaxation time is just $t$,
\begin{equation}
  \label{eq:transverse-corr-length}
  1=\lpe^{-2}\left[\lpe^{-2} t +F(t)\right] \;.
\end{equation}
Asymptotically $\lpe(t)$ is given by
\begin{equation}
  \label{eq:lpe-asymptotically}
  \lpe(t)\sim\left\{ {t^{1/4}\;, \text{ for }t\ll (F/t)^{-2}
      \atop F^{1/2} \;, \text{ for }t\gg
      (F/t)^{-2}}\right. \;.
\end{equation}

\begin{table}
  \caption{The transverse equilibration length
    $\lpe(t)$ and the tension propagation length
    $\lpa(t)$ both exhibit a crossover at a time $t_\fe\equiv\fe^{-2}$, which depends on
    the external force $\fe$ 
    (here, for the {\pull} problem with $\fe\gg L^{-2}$).}
  \label{tab:pulling-growth-laws}
  \begin{center}
    \begin{tabular}{r|cc|c} &
      $\lpe(t)$  & & $\lpa(t)$ \qquad \quad \mbox{}\\ \hline
      $t\ll t_\fe$ &  $t^{1/4}$ &  & $t^{1/8} (\lp/ \zeta )^{1/2}$ 
      \hfill~\cite{morse:98II,everaers-Maggs:99}\\
      $t\gg t_\fe$ &  $t^{1/2}\fe^{1/2}$ & & $t^{1/4}\fe^{1/4} 
      (\lp/\zeta)^{1/2}$ \quad ~\cite{seifert-wintz-nelson:96}
    \end{tabular}
  \end{center}
\end{table}

Due to the competition between bending forces ($\propto r_\perp\lpe^{-4}$) and
tension ($\propto r_\perp (F/t) \lpe^{-2}$), the growth of $\lpe$ thus exhibits a
dynamic crossover from free relaxation ($\lpe\simeq t^{1/4}$) to relaxation under
tension ($\lpe\simeq \sqrt{F}$) at a characteristic time $t_\fe\equiv (F/t)^{-2}$
(Table~\ref{tab:pulling-growth-laws}/left for a constant tension equal to the
external force $\fe$).

As we will explicitely demonstrate in the next section, the change $\avg{\Delta
  \varrho_{\frac hc}}$ in stored length saturates at a constant value for distances
to the boundaries much larger than the characteristic length $\lpe(t)$. This ``bulk''
value is given by~\footnote{This equation for the stored length at a spatially
  constant but time-dependent tension has recently independently been derived in a
  related context~\cite{bhobot-wiggins-granek:04}.}
\begin{eqnarray}
    \avg{\Delta\overline{\varrho}}(t)&\equiv& \mint{\frac{dq}{\pi\lp}}{0}{\infty}
    \left\{\frac{1}{q^2+f_<}\left(e^{-2q^2[q^2 t+ F(t)]}-1\right) \right.\nonumber\\
    &&+\left.2 \theta
      q^2\mint{d\tilde t}{0}{t}e^{-2q^2\left[q^2(t-\tilde
          t)+F(t)-F(\tilde t)\right]}\right\}. \label{eq:change-stored-length}
\end{eqnarray}
The quantity $\avg{\Delta\overline{\varrho}}(t)$ will be central in our systematic
analysis of tension propagation and relaxation, because it turns out to determine the
local curvature of the tension profile.  Each of the two terms inside the curly
brackets of Eq.~(\ref{eq:change-stored-length}) have a direct physical
interpretation.  Since the parameter $\theta$ tunes the strength of the thermal
kicks, it is seen that the $\theta$-independent first term represents the
deterministic change in the excess length that is stored in mode $q$ in absence of
any stochastic force.  For pulling forces $F>0$ its sign is always negative, since
both the internal elastic and the external driving forces act to straighten the
filament.  On the contrary, thermal kicks represented by the (strictly positive)
second term are favoring undulations.

\section{Breakdown of ordinary perturbation theory}
\label{sec:intro-tensprop-sfp}
The previous sections employed ``ordinary'' perturbation theory (OPT) in the small
parameter $\epsilon$, leading to a linear equation of motion to lowest order. As
detailed below, the use of OPT is, however, limited to long times even for
$\epsilon\ll1$.  The predictions derived above for the longitudinal segment motion
turn out to be incompatible with the longitudinal force balance on short times. In
particular, Eq.~(\ref{eq:change-stored-length}) reveals an infinite longitudinal
friction for $t\to0$. This section extends a heuristic argument of
Ref.~\cite{everaers-Maggs:99} to resolve this problem.  The following, furthermore,
elucidates a very general feature of the non-linear response, namely, a crossover
from ``weak-'' to ``strong-force'' behavior. Finally, it reveals the crucial
length-scale separation underlying our subsequent systematic analysis.

\begin{figure}
  \includegraphics[width=\columnwidth]
  {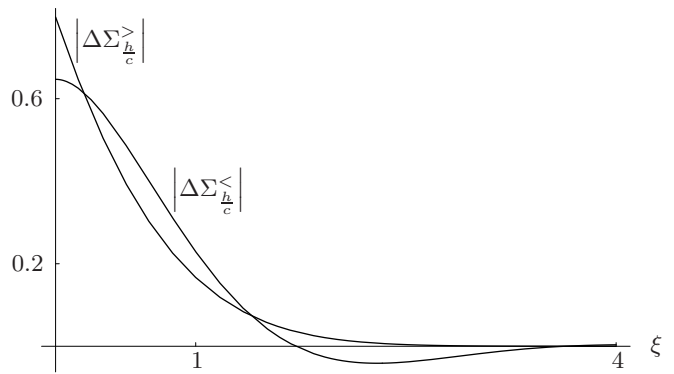}
  \caption{The absolute value $\abs{\Delta\Sigma^{\gtrless}_{\frac
        hc}}$ of the spatially varying part $\Delta\Sigma^{\gtrless}_{\frac
      hc}\equiv\Sigma^{\gtrless}_{\frac hc}(\xi)-\Sigma^{\gtrless}_{\frac hc}
      (\infty)$ of the scaling functions $\Sigma_{\frac hc}^\gtrless$ defined in
        Eqs.~(\ref{eq:drho-pull-3}, \ref{eq:drho-pull-4}), which happens
      to be the same for hinged/clamped boundary conditions.}
  \label{fig:blayer-pull-opt}
\end{figure}

The breakdown of OPT becomes evident when we try to use
Eq.~(\ref{eq:delta-rho-mega-expression}) to evaluate the longitudinal segment motion
in a non-equilibrium situation. For definiteness and as a telling example, let us
consider an initially equilibrated polymer that is suddenly pulled longitudinally by
a constant force $\fe$ at both ends, i.e., at time $t=0$, the tension at the ends is
suddenly increased from $0$ to a given positive value $\fe$,\[\fe=f(0,t>0)=f(L,t>0)
\quad\text{({\pull}-scenario).}\] As a consequence of Eq.~(\ref{eq:eompa-low}), the
tension $f=\fe \Theta(t)$ is to lowest order spatially uniform and fixed by the
driving force $\fe$ at the boundaries. Our above result for the change in stored
length due to a given tension history, Eq.~(\ref{eq:delta-rho-mega-expression}), thus
applies and evaluates to
\begin{eqnarray}
  \label{eq:drho-pull-1}
  \avg{\Delta\varrho_{\frac hc}}(s,t)&=&\mint{\frac{dq}{\pi\lp}}{0}{\infty}
  \frac{\fe}{q^2(q^2+\fe)}\left[ e^{-2q^2(q^2+\fe)t} -1 \right]\nonumber\\ 
  &&\times\left[1\pm\cos(2 q s)\right]   \;.
\end{eqnarray}
After the variable substitutions $k\equiv q\fe^{-1/2}$, $\sigma\equiv s \fe^{1/2}$ and
$\tau\equiv \fe^2 t=t/\tf$, Eq.~(\ref{eq:drho-pull-1}) takes the form
\begin{equation}
  \label{eq:scform-1}
  \avg{\Delta
    \varrho_{\frac hc}}(s,t)=\lp^{-1}\fe^{-1/2}\Sigma_{\frac hc}(\sigma,\tau)
\end{equation}
with the
scaling function
\begin{eqnarray}
  \label{eq:drho-pull-2}
  \Sigma_{\frac hc}(\sigma,\tau)&=&\mint{\frac{dk}{\pi}}{0}{\infty}
  \frac{1}{k^2(k^2+1)}\left[ e^{-2k^2(k^2+1)\tau} -1 \right]\nonumber \\
  & &\times\left[1\pm\cos(2 k \sigma)\right]   \;.
\end{eqnarray}
Another variable substitution $\tilde k=k \tau^{1/4}$ generates factors $\tilde
k^4+\tilde k^2\tau^{1/2}$ in exponent and denominator, which can be replaced by
$\tilde k^4$ in the asymptotic limit $\tau\ll 1$. Just as for the dispersion relation
Eq.~(\ref{eq:relaxation-times}), tensile forces $\propto k^2$ may be neglected as
compared to the dominant bending forces $\propto k^4$ for times smaller than the
crossover time $\tf$. In the opposite limit $\tau\gg 1$, the reverse approximation
applies. After a variable substitution $\tilde k=k \tau^{1/2}$, factors $\tilde k^4
\tau^{-1}+\tilde k^2$ appear, which may be replaced by $\tilde k^2$. We thus find
that the two-parameter scaling form Eq.~(\ref{eq:drho-pull-2}) collapses onto
one-parameter scaling forms for small and large times,
\begin{eqnarray}
  \label{eq:scf-stimes}
  \Sigma_{\frac hc}&\sim& -\tau^{3/4}\Sigma_{\frac hc}^<(\sigma
  \tau^{-1/4})\;,\qquad \tau\to 0 \nonumber\\
  \label{eq:scf-ltimes}
  \Sigma_{\frac hc}&\sim& -\tau^{1/2}\Sigma_{\frac hc}^>(\sigma
  \tau^{-1/2})\;,\qquad \tau\to \infty \nonumber
\end{eqnarray}
with scaling functions given by
\begin{eqnarray}
  \label{eq:drho-pull-3}
  \Sigma_{\frac hc}^<(\xi)&=&\mint{\frac{d\tilde k}{\pi}}{0}{\infty}\frac{1-e^{-2 \tilde
      k^4}}{\tilde k^4}\left[ 1\pm\cos(2 \tilde k \xi)  \right]\;,\\
  \label{eq:drho-pull-4}
  \Sigma_{\frac hc}^>(\xi)&=&\mint{\frac{d\tilde k}{\pi}}{0}{\infty}\frac{1-e^{-2 \tilde
      k^2}}{\tilde k^2}\left[ 1\pm\cos(2 \tilde k \xi)  \right]\;.
\end{eqnarray}
Note that the spatial part of these scaling functions depicted in
Fig.~\ref{fig:blayer-pull-opt} decays to zero within several rescaled time units.  As
an important consequence, we note that the part of $\avg{\Delta \varrho_\frac hc}$
that depends on the boundary conditions really only matters close to the boundaries,
i.e., up to a distance for which the scaling variable becomes of order one. In fact,
this distance can be identified with $\lpe(t)$, as defined in
Eq.~(\ref{eq:transverse-corr-length}), because the scaling variable of $\Sigma$ is
given by $\sigma \tau^{-1/4}=s/t^{1/4}=s/\lpe(t)$ for $\tau\ll1$ and $\sigma
\tau^{-1/2}=s/\lpe(t)$ for $\tau\gg1$, respectively.

The bulk of the polymer stores length according to the universal part
$\avg{\Delta\overline{\varrho}}(t)$ (independent of the boundary condition), which
asymptotically takes the form
\begin{equation}
  \label{eq:DR-opt}
  \avg{\Delta \overline\varrho}(t)\sim \left\{ { \Gamma(7/4)^{-1} \fe
      \lp^{-1}   (t/2)^{3/4}
     \text{~\cite{granek:97}}    \;,\quad \text{for } t\ll \tf  \atop 
       \lp^{-1} (2\fe t/\pi)^{1/2} \;, \quad \text{for } t\gg \tf}
  \right. \;.
\end{equation}
For alternative derivations of the short-time linear response law $\sim t^{3/4}$, see
Refs.~\cite{granek:97,gittes-mackintosh:98_pub,morse:98II}. The crossover time
$\tf\equiv \fe^{-2}$ is the time where the external force $\fe$ equals the Euler
buckling force $\lpe^{-2}(t)$ corresponding to the correlation length $\lpe(t)$.

These results comprise the predictions of ordinary perturbation theory (OPT) to
leading order. As evident from Eq.~(\ref{eq:eom1}), longitudinal friction forces have
thereby been completely neglected, because they are of higher order in $\epsilon$.
However, on the semi-infinite arclength interval considered here, a spatially
constant change in stored length - no matter how small - implies via
Eq.~(\ref{eq:end-end-length}) an infinitely fast change $\partial_t \avg{\Delta
  R_\pa}$ of the longitudinal extension and thus friction. Hence, OPT must fail on
sufficiently large length scales. Let us define $L_\star(t)$ as the length scale
beyond which OPT breaks down for a given time $t$.  In the present case of a suddenly
applied pulling force, this critical length can be estimated from the physical
requirement that the total longitudinal friction is not only finite but at most equal
to the driving force $\fe$, i.e.,
\begin{equation}
  \label{eq:condition}
  \fe\stackrel{!}{\simeq} \hat \zeta L \partial_t
  \avg{\Delta R_\pa}=-\hat \zeta L\mint{ds}{0}{L}\partial_t\avg{\Delta\varrho}(s,t)\;.
\end{equation}
For an average stored length given by the spatially
constant value Eq.~(\ref{eq:DR-opt}), this condition is met if the polymer length is
smaller than
\begin{eqnarray}
  \label{eq:critical-length}
  L_\star(t) &\equiv&\sqrt{\fe/(\hat \zeta\partial_t\avg{\Delta\overline{\varrho}})} \nonumber\\
  &\simeq&\left\{{ \sqrt{\lp/\hat \zeta}\;t^{1/8}  \;,\qquad \text{for } t\ll \tf  \atop 
      \sqrt{\lp /\hat \zeta}\;(\fe t)^{1/4}  \;, \qquad \text{for } t\gg \tf}\right.\;.
\end{eqnarray}
OPT can thus only be valid on length scales (much) smaller than $L_\star(t)$. The
condition
\begin{equation}
  \label{eq:tstar}
  L_\star(t_\star)=L \;,
\end{equation}
implicitly defines the critical time scale $t_\star$ above which OPT applies to the
whole polymer, and below which OPT is limited to subsections shorter than
$L_\star(t)$.

In summary, the omission of tension propagation has been identified as the reason for
the breakdown of the OPT predictions. On a heuristic level~\cite{everaers-Maggs:99},
the problem can thus be resolved by requiring that the applied tension is not
immediately perceptible everywhere in the filament, but propagates a finite distance
$\lpa(t)$ during time $t$ from the ends towards the bulk of the filament. Hence, only
segments up to a distance $\lpa(t)$ from the ends are set into longitudinal motion.
If the length $\lpa(t)$ over which the tension varies is smaller than the critical
length $L_\star(t)$, it is ensured that the longitudinal friction does not exceed the
driving force.  In heuristic discussions, it was generally assumed that both lengths
can be identified up to  numerical factors of order one,
\begin{equation}
  \label{eq:heuristic-argument}
  \lpa(t)\simeq L_\star(t)\;,\;t\ll t_\star\;,\quad \text{(heuristic hypothesis)}
\end{equation}
as summarized in Table \ref{tab:pulling-growth-laws}/right. The scaling assumption
in Eq.~(\ref{eq:heuristic-argument}) turns out to reproduce the correct scaling for most
cases (with, however, interesting exceptions elaborated in Part~II~\cite{hallatschek-part2:2006}, as well as, in
Ref.~\cite{obermayer-kroy-frey-hallatschek3:tbp}).  The ``weak- and strong- force''
limits $\lpa\propto t^{1/8}$ and $\lpa\propto t^{1/4}$ of
Refs.~\cite{everaers-Maggs:99,seifert-wintz-nelson:96} are with
Eqs.~(\ref{eq:critical-length},~\ref{eq:heuristic-argument}) recovered as ``short- and
long-time'' asymptotics. The crossover at time $\tf$ signals the change from ``free''
to ``forced'' relaxation and is inherited from the one of the scaling function
$\Sigma$, defined in Eq.~(\ref{eq:drho-pull-2}).

Finally, we note that the tension may be considered as ``slowly''
varying in space because $\lpa(t) \propto \Ord{\epsilon^{-1/2}}$ is
for $\epsilon\to0$ larger than any length that does not dependent on
the small parameter (for $t,$ $\fe,$ $L$ fixed).  As it turns
out, the most important length of the latter type is the dynamic
correlation length $\lpe(t)$ for transverse
displacements. Namely, the \emph{scale separation}
\begin{equation}
  \label{eq:lscale-sep}
  \lpa/\lpe=\Ord{\epsilon^{-1/2}} \gg 1\;
\end{equation}
indicates that the tension is nearly constant on the equilibration length scale
$\lpe(t)$ for transverse displacements.  Intuitively, it should be clear that this
simplifies the further analysis considerably, because it allows to apply (certain)
results locally that are derived for spatially constant tension.  Formally,
Eq.~(\ref{eq:lscale-sep}) lends itself as a starting point for a \emph{multiple-scale}
calculus, which separates the physics on different scales to obtain an improved
perturbation expansion that is regular in the limit $t\to0$ while $\epsilon\ll1$ is
fixed~\footnote{It may be remarked that the encountered contradiction does not
  appear, if one takes the limit $\epsilon\to0$ while the parameters $L$ and $t$ are
  held fixed.  This corresponds to a lower temperature, respectively, a stiffer
  polymer.  Then, the physically motivated requirement of the external force
  exceeding the total longitudinal friction force is satisfied for small enough
  $\epsilon\ll1$.  The same conclusion may be drawn from the cross-over length scales
  $L_\star$ in Eq.~(\ref{eq:critical-length}) upon eliminating $\lp$ in favor of the
  small parameter $\epsilon$, because $L_\star(\epsilon,t) >L$ for given $L$ and $t$
  and small enough $\epsilon$.  In a mathematical sense, the expansion generated by
  ordinary perturbation is pointwise asymptotic in $t$, but not
  uniformly~\cite{hinch:91}: the smaller $t$ the smaller $\epsilon$ has to be for the
  expansion to be asymptotic.}.  The procedure, detailed in the next section, is
similar in spirit to the procedure for athermal rod
dynamics~\cite{hallatschek-frey-kroy:04}, but some complications related to the
stochastic nature of the equations of motion have to be faced. The final result will
be an effective deterministic description of the tension on the macro-scale
$\lpa(t)$, where the stochasticity on the micro-scale $\lpe(t)$ has been integrated
out.

\section{Multiple scale analysis}
\label{sec:multi-scale}
We introduce a rapidly and a slowly varying arc length coordinate, $x\equiv s$ and
$y\equiv s \epsilon^{\gamma}$, respectively, where the exponent $\gamma>0$ will be
fixed later. The dynamic functions $\vec r_\pe$ and $f$ are now considered to depend
on both variables $\{f,\vec r_\pe\}\to \{f(x,y),\vec r_\pe(x,y)\}$, where $x$ and $y$
are treated as independent.  The original arc length derivative of those functions
then becomes
\begin{equation}
  \label{eq:spatio-derivative}
  \partial_s \equiv \partial_x|_{y} + \epsilon^\gamma
  \partial_y|_{x} \;.
\end{equation}
The noise $\vec \xi=\Ord{\epsilon^{1/2}}$ being the source of any
transverse displacements suggests an expansions of the dynamic
variables $\vec r_\perp$ and $f$ in powers of $\epsilon^{1/2}$,
\begin{eqnarray}
  \label{eq:power-series}
  \vec r_\perp(x,y)&=&\epsilon^{1/2} \vec h_1(x,y)+ o\left(\epsilon^{1/2}\right) 
  \;, \nonumber \\
  f(x,y) &=& f_0(x,y)+\epsilon^{1/2} f_1(x,y)+\epsilon f_2(x,y)\nonumber \\
  & &+o(\epsilon) \;.
\end{eqnarray}
In the case of isotropic friction (i.e.~$\zeta_\pe=\zeta_\pa$), the stochastic forces
have no intrinsic scale. Hence, they can only depend on the microscopic variable,
$\vec \xi=\epsilon^{1/2}\vec \xi_1(x)$. In the anisotropic case, the friction forces
and, hence, the stochastic forces, are coupled to the orientation of the filament, so
that one has to assume a power expansion
\begin{equation}
   \label{eq:noise-expansion}
  \begin{split}
    \boldsymbol \xi_\pe(x,y)&=\epsilon^{1/2}\boldsymbol
    \xi_{\pe,1}(x)+o\left(\epsilon^{1/2}\right)  \\
    \xi_\pa(x,y)&=\epsilon^{1/2} \xi_{\pa,1}(x)+\epsilon \xi_{\pa,2}(x,y)+o(\epsilon)
    \;.
 \end{split}
\end{equation}
The $y$-arguments in Eq.~(\ref{eq:noise-expansion}) are inherited from the
$y$-arguments of $\vec r'$ entering the noise correlations in
Eq.~(\ref{eq:noise-corrs}) via the friction matrix, defined in
Eq.~(\ref{eq:friction-tensor}). The $y$-dependence would disappear for isotropic
friction. Note, that the leading order $\vec \xi_1(x)$ still depends on the
microscopic variable $x$ only, because the anisotropy merely enters the higher orders
(a formal argument is given in App.~\ref{sec:noise-mspt}).

In the following, it is crucial to require that the expansion coefficients in each
order are bounded, so that we obtain a uniformly valid power
expansion~\cite{holmes:95} in terms of the small parameter $\epsilon$.  Inserting all
expansions, Eqs.~(\ref{eq:noise-expansion},~\ref{eq:power-series}), into the
equations of motion, Eq.~(\ref{eq:perp-eq-motion-sfp},
\ref{eq:parallel-eq-motion-sfp}), yields
\begin{subequations}
  \begin{align}
    0&= \epsilon^{1/2}\left[\partial_t \vec h_1+\partial_x^4\vec
      h_1-\partial_x(f_0 \partial_x \vec
      h_1)-\boldsymbol \xi_{\pe,1}\right] \nonumber \\
    &\quad + o(\epsilon^{1/2}) \label{eq:multiscale-1} \\
    0&=\partial_x^2 f_0 +\epsilon^\gamma 2 \partial_x\partial_y f_0
    +\epsilon^{1/2} \left[\partial_x^2
      f_1+\xi_{\pa,1}'\right]\nonumber\\
    &\quad +\epsilon^{2\gamma}\partial_y^2 f_0
    +\epsilon\left[\partial_x^2 f_2+X_2(x,y)+\xi_{\pa,2}'\right]
    \nonumber \\
    &\quad +\epsilon^{\gamma+1/2}\partial_y\partial_x f_1+
    o(\epsilon;~\epsilon^{2\gamma}) \;.
    \label{eq:multiscale-2}
  \end{align}
\end{subequations}
In order to arrive at Eq.~(\ref{eq:multiscale-2}) we used the local arc length
constraint, Eq.~(\ref{eq:constraint-wbl-sfp}).  By
\begin{eqnarray} 
  \label{eq:nl}
  X_2(x,y)&=&\frac{\hat \zeta}{2}\partial_t(\partial_x \vec h_1)^2+(1-\hat
  \zeta)\partial_x\left[(\partial_x \vec h_1) (\partial_t\vec
    h_1)\right]\nonumber\\
  &&+\frac{1}{2}\partial_x^4\left(\partial_x \vec
    h_1\right)^2-\frac{1}{2}\partial_x^2\left[f_0\left(\partial_x\vec
      h_1\right)^2  \right]
\end{eqnarray}
we have summarized terms nonlinear in $\vec h_1$. The first term in $X_2$, which is
proportional to $ \hat \zeta$, accounts for the longitudinal friction and is thus
responsible for the short-time divergence encountered in the heuristic discussion of
Sec.~\ref{sec:intro-tensprop-sfp}. 

The $\Ord{1}$ part of Eq.~(\ref{eq:multiscale-2}), $\partial_x^2
f_0=0$, together with the requirement of $f_0$ being bounded for large
$x$, implies that
\begin{equation}
  \label{eq:x-indep-tension}
  f_0(x,y)=\hat f_0(y)
\end{equation}
is independent of $x$. Hence both the $\Ord{1}$ and the $\Ord{\epsilon^\gamma}$ term
of Eq.~(\ref{eq:multiscale-2}) vanish.  Requiring the $\Ord{\epsilon^{1/2}}$
coefficient to be zero fixes $f_1$ up to an integration constant. The value of $f_1$
does not affect the evolution of $\vec h_1$ because it does not enter the
$\Ord{\epsilon}$ coefficient in Eq.~(\ref{eq:multiscale-1}).  Thus, the precise value
of $f_1$ does not change the pathological behavior of longitudinal friction $\propto
\hat \zeta \partial_t (\partial_x \vec h_1)^2$, which is why we shift the discussion
of $f_1$ to App.~\ref{sec:f1}.  From the latter we merely need that
$\partial_y\partial_x f_1(x,y)=0$ which renders the
$\Ord{\epsilon^{\gamma+1/2}}$-term in Eq.~(\ref{eq:multiscale-2}) zero. Then the next
higher order is either $\Ord{\epsilon^{2\gamma}}$ or $\Ord{\epsilon}$ depending on
the value of $\gamma$. With Eq.~(\ref{eq:x-indep-tension}) we can solve the
$\Ord{\epsilon^{1/2}}$ part of Eq.~(\ref{eq:multiscale-1}) for $\vec h_1(x,y)$ along
the lines of Sec.~\ref{cha:LT} and use the result to evaluate $X_2(x,y)$.  It then
turns out that the first term in $X_2$ (the longitudinal friction) would require
$f_2$ to grow without bound with increasing system size to render the
$\Ord{\epsilon}$-expansion coefficient in Eq.~(\ref{eq:multiscale-2}) finite, if they
were required to balance each other. This represents the same unphysical divergence
that is responsible for the breakdown of ordinary perturbation theory, discussed in
Sec.~\ref{sec:intro-tensprop-sfp}.

In order to obtain an improved perturbation theory, we attempt to balance the
nonlinear term by the $\Ord{\epsilon^{2\gamma}}$ term after choosing $\gamma=1/2$;
i.e., the exponent $\gamma$ is fixed such that the expansion coefficient $f_2$
remains finite in the semi-infinite system considered here~\footnote{We note aside
  that the small parameter $\epsilon^\gamma=\epsilon^{1/2}$ appearing here is the
  same as in the length scale separation, Eq.~(\ref{eq:lscale-sep}), observed in
  Sec.~\ref{sec:intro-tensprop-sfp}.}.  The equation fixing $f_2$ thus reads
\begin{eqnarray}
  \label{eq:phi1}
  \partial_x^2f_2(x,y)&=&-\partial_y^2\hat
  f_0(y)-X_2(x,y)-\xi_{\pa,2}'\;.
\end{eqnarray}
Given $h_1$ and $f_0$ the last equation can be solved for $f_2$
\begin{eqnarray}
  \label{eq:fi-2}
  f_2(x,y)&=&\mint{d\tilde x}{0}{x}\mint{d\hat x}{0}{\tilde
    x}\left\{ -\partial_y^2\hat
    f_0(y)\right.\nonumber\\
  &&\left.-X_2(\hat x,y)-\xi_{\pa,2}'(\hat x) \right\}  \,.
\end{eqnarray}
For $f_2$ to be bounded for large system sizes, we have to require
\begin{equation}
  \label{eq:fixing-f0}
  \partial_y^2 \hat
  f_0(y)=\overline{-X_2-\xi_{\pa,2}'\,}^x(y) \;,
\end{equation}
where the over-line denotes the spatial average over the rapidly
varying coordinate $x$,
\begin{equation}
  \label{eq:xavg} \overline{g}^x(y)\equiv \lim_{l\to\infty}
  \mint{\frac{dx}{l}}{0}{l}g(x,y)
\end{equation}
for a function $g(x,y)$. The expansion coefficient $f_2$ would show a divergence
quadratic in the system size if Eq.~(\ref{eq:fixing-f0}) was not satisfied. Hence,
the $y$-dependence of $\hat f_0(y)$ must be fixed such that the expansion coefficient
$f_2$ remains finite. For a finite polymer, the limit $l\to\infty$ is not to be taken
literally, though. Rather, the average in Eq.~(\ref{eq:xavg}) is required to become
independent of $l$ to leading order in $\epsilon$ for $l$ much smaller than the
system size $L$.

As it turns out, the only quantity in Eq.~(\ref{eq:fixing-f0}) that
does not vanish upon $x$-averaging is the first term in
$X_2$, the longitudinal friction.  This is easily
seen for all other terms in $X_2$ and the $f_1$-term. They are total
derivatives with respect to $x$ of products of expansion coefficients
that are (required to remain) bounded (non-secular~\cite{holmes:95})
by definition.  Hence, the $x$-integrals of those total derivatives
are bounded and the $x$-averages vanish upon formally taking the
coarse-graining length $l\to\infty$ in Eq.~(\ref{eq:xavg}).  The
noise term also represents a total derivative with respect to $x$. The
average of that term represents a stochastic variable with an
amplitude that scales as $1/l$ and, hence, also vanishes in the limit
$l\to\infty$.

Dropping all terms that vanish under coarse-graining and integrating
over time, Eq.~(\ref{eq:fixing-f0}) takes the form
\begin{eqnarray}
  \label{eq:cg-eom1}
  \frac1{\hat\zeta}\partial_y^2\hat F_0(y)&=&\frac12 \left
    [\overline{(\partial_x \vec h_1)^2}^x(y,0)-\overline{(\partial_x
      \vec h_1)^2}^x(y,t)\right] \nonumber \\
  &=&-\epsilon^{-1}\overline{\Delta\varrho}^s(t)\;,
\end{eqnarray}
where $\Delta \varrho(s,t)$ is the change in stored length, as defined in
Eq.~(\ref{eq:change-in-slength}), of a semi-infinite polymer for the tension history
\begin{equation}
  \label{eq:tension-history-msp}
  F(t)=\hat F_0(y,t) \;.
\end{equation}
Note that the dependence on the slowly varying arclength coordinate $y$ enters the
tension history in Eq.~(\ref{eq:tension-history-msp}) only parametrically. The same
holds true for the calculation of the right hand side of Eq.~(\ref{eq:cg-eom1}).

The closed set of equations Eqs.~(\ref{eq:cg-eom1},~\ref{eq:multiscale-1}) represents
the lowest order of the multiple scale perturbation expansion. It incorporates the
feedback mechanism already found in the heuristic discussion above.  The evolution
of transverse displacements implies longitudinal motion via the arc length
constraint, and, according to Eq.~(\ref{eq:cg-eom1}), the corresponding longitudinal
friction sets the polymer under tension. This, in turn, feeds back onto the evolution
of transverse displacements, Eq.~(\ref{eq:multiscale-1}), typically acting as to
reduce the longitudinal friction.

For solving Eqs.~(\ref{eq:cg-eom1},~\ref{eq:multiscale-1}) self-consistently, it
would be handy to perform an ensemble average $\avg{\dots}$ on the right-hand-side
of Eq.~(\ref{eq:cg-eom1}), because we could then apply expression
Eq.~(\ref{eq:delta-rho-mega-expression}) for $\avg{\Delta\varrho}$. We argue that
such an ensemble average is indeed justified, because $\overline{\Delta\varrho}^s(t)$
is, in fact, a deterministic quantity as a consequence of the central limit theorem.
Recall that the quantity $\avg{\Delta{\varrho}}(s,t)$ is dominated by transverse
modes of the wave length $\lpe(t)$ given by Eq.~(\ref{eq:lpe-asymptotically}). Since
wavelength much larger than $\lpe(t)$ are not relevant in the mode sum, the dynamic
length $\lpe(t)$ can be interpreted as the correlation length of the change in stored
length, i.e., the length over which the correlation function
$C_{s,t}(z)\equiv\avg{\Delta\varrho(s,t)\Delta\varrho(s+z,t)}$ varies.  Hence, the
integral $X_l\equiv \mint{ds}{0}{l}\Delta\varrho(s,t)$ may be understood as the sum
of $l/\lpe(t)$ weakly correlated random variables. For large $l$, the distribution
function of $X_l$ thus becomes Gaussian, and the variance of $X_l$ grows linearly
with the number of independent random variables, $\avg{(X_l-\avg{X_l})^2}\propto l$.
As a consequence, the distribution of the average $X_l/l\to
\overline{\Delta\varrho}^s$ approaches a delta function as $l\to\infty$.

An ``additional'' ensemble average therefore does not change the value of the right
hand side of Eq.~(\ref{eq:cg-eom1}). Evaluating the spatial average
after the ensemble average levels out the boundary term of expression
Eq.~(\ref{eq:delta-rho-mega-expression}) for $\avg{\Delta \varrho_{\frac hc}}$ and
reduces it to its \emph{bulk} value $\avg{\Delta\overline{\varrho}}(t)$, defined in
Eq.~(\ref{eq:change-stored-length}), which is completely independent of the boundary
conditions. We thus have the important relation
\begin{equation}
  \label{eq:self-averaging}
  \overline{\Delta\varrho_{\frac hc}}^s(t)=\overline{\avg{\Delta\varrho_{\frac
      hc}}}^s(t)=\avg{\Delta\overline{\varrho}}(t) \;.
\end{equation}
We would like to emphasize, that our argument for replacing the spatial by an
ensemble average requires a finite driving force, such that
$\lim_{l\to\infty}\overline{\Delta \varrho}^s$ approaches a finite value. This
specifically excludes the linear response limit, i.e., the limit of vanishing
external force $\fe\to0$ while $\epsilon\ll1$ is fixed. In this case, the tension
dynamics has to be described by Eq.~(\ref{eq:cg-eom1}), which is \emph{stochastic}
even to leading order (we will come back to this point in Part~II).

Given the external driving is finite such that Eq.~(\ref{eq:self-averaging}) may be
applied, Eq.~(\ref{eq:cg-eom1}) takes the form
\begin{equation}
  \label{eq:cg-eom}
  \partial_s^2 F(s,t)=-\hat \zeta \avg{\Delta \overline{\varrho}}\left[F(s,\tilde t
    \leq t),t\right]  \;,
\end{equation}
where we introduced $s=y\epsilon^{1/2}=x$ again and made the
parametric dependence of $\avg{\Delta\overline{\varrho}}$ on the
tension history explicit. The \emph{deterministic} tension dynamics,
as described by Eq.~(\ref{eq:cg-eom}), provides the sought-after
rigorous local generalization of the heuristic argumentation of
Sec.~\ref{sec:stored-length}: local longitudinal motion is driven by
tension gradients (like in a thread pulled through a viscous medium).

Upon inserting our result in Eq.~(\ref{eq:change-stored-length}) for the
right-hand-side, we obtain a nonlinear, partial integro-differential equation (PIDE)
for the tension history $F(s,t)$,
\begin{eqnarray}
  \label{eq:pide}
    &\partial_s^2 F(s,t)= \hat \zeta
    \mint{\frac{dq}{\pi\lp}}{0}{\infty}
    \left\{\frac{1}{q^2+f_<}\left(1-e^{-2q^2[q^2 t+ F(s,t)]}\right)
    \right.& \nonumber \\
    &\left. -2 \theta q^2\mint{d\tilde
        t}{0}{t}e^{-2q^2\left[q^2(t-\tilde t)+F(s,t)-F(s,\tilde
          t)\right]}\right\}\;.&
\end{eqnarray}
We have arrived at a closed description of the polymer dynamics to lowest
order in MSPT that consists of two parts. On a length scale $\lpe(t)$, Brownian
motion gives rise to fluctuations of transverse displacements that are described by
the linear Eq.~(\ref{eq:multiscale-1}). This \emph{stochastic} differential
equation, in turn, adiabatically depends on a tension profile that varies on a much
larger scale $\lpa(t)$ and satisfies a \emph{deterministic} nonlinear equation of
motion, Eq.~(\ref{eq:pide}).

\begin{figure}
    \centerline{\includegraphics[width=\columnwidth] {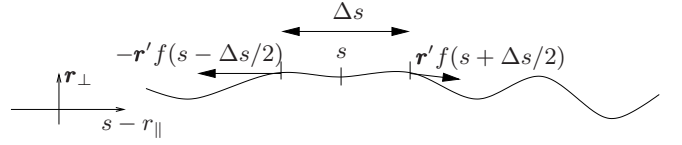}}
  \caption{Tensile forces acting on a polymer subsection of size $\Delta
    s$.  Balancing these forces with the drag that  arises from the longitudinal velocity
    of magnitude $\partial_t r_\pa(s,t)$, one estimates $\Delta s \zeta_\pa \partial_t
    r_\pa(s,t)\approx f(s-\Delta s/2)-f(s+\Delta s/2)$ in the weakly bending limit.
    For infinitesimal $\Delta s$, this becomes $\zeta_\pa \partial_t
    r_\pa(s,t)\approx-\partial_s f(s,t)$.  A further spatial derivative yields
    $\zeta_\pa \partial_t\partial_s r_\pa(s,t)\approx-\partial_s^2 f(s,t)$, which is
    the time derivative of Eq.~(\ref{eq:cg-eom}) (in original units) up to a spatial
    and ensemble average on the left-hand side, which correspond to an adiabatic,
    respectively, equilibrium approximation. }
  \label{fig:tension-vs-friction}
\end{figure}

\section{Conclusion}
\label{sec:conclusion1}

We would like to conclude the present general discussion of tension dynamics with a
simple physical interpretation of the outcome, Eq.~(\ref{eq:cg-eom},~\ref{eq:pide}),
of our multiple scale analysis. Effectively, our MSPT analysis is a rigorous
justification of certain approximations that can be made to analyze the tension
dynamics in the small $\epsilon$ limit. According to Eq.~(\ref{eq:cg-eom}), the
curvature of the integrated tension is (up to a constant) given by the ensemble
average of the local stored length release. As shown in
Fig.~\ref{fig:tension-vs-friction}, it may be conceived as (1) a force balance
equation between the locally acting tensile and longitudinal friction force, in which
(2) the latter may be computed from the equations of motion for the bulk of an
equilibrated polymer under a spatially constant, though time-dependent, tension. The
longitudinal friction force is obtained via taking a time-derivative of this
coarse-grained \emph{dynamical} force extension relation,
Eq.~(\ref{eq:change-stored-length}). The first order MSPT equation of motion neglects
longitudinally acting bending forces, employs an adiabatic approximation and assumes
local equilibrium. The latter fully retains memory effects and therefore must not be
mistaken as an approximation of quasi-stationary dynamics, which, in
Part~II~\cite{hallatschek-part2:2006}, will turn out to be a valid approximation only
for specific driving forces in a particular time regime.

\section{Acknowledgments}
\label{sec:ack}

It is a pleasure to acknowledge helpful conversations with Benedikt Obermayer. This
research was supported by the German Academic Research Foundation (DAAD) through a
fellowship within the Postdoc-Program (OH) and by the Deutsche
Forschungsgemeinschaft through grant no.~Ha 5163/1 (OH) and SFB 486 (EF).

\appendix

\begin{table}[b]
  \caption{Some important notations}
  \label{tab:common-notation}
  \begin{ruledtabular}
    \begin{tabular}{c|l}
      Symbol(s) & General Meaning \\ \hline
      $L$ & total length of the worm-like chain \\
      $\kappa$ & bending stiffness \\
      $\epsilon$  & small parameter, defined such that $\vec r_\pe'^2=\Ord{\epsilon}$    \\
      $\vec r_\pe(s,t)$ & transverse displacement; $\vec r_\pe=\Ord{\epsilon^{1/2}}$ \\
      $r_\pa(s,t)$ & longitudinal displacement; $\vec r_\pa=\Ord{\epsilon}$ \\
      $\varrho(s,t)$ & stored-length density; $\varrho=\vec
      r_\pe'^2/2+\Ord{\epsilon^2}=\Ord{\epsilon}$  \\
      $R$ & end-to-end distance \\
      $R_\pa$ & end-to-end vector projected onto the long.~axis
      \\
      $\simeq$ & equal up to numerical factors of
      order $1$ \\
      $\propto$ & proportional to \\
      $\sim$ & asymptotically equal \\
      $\lp$ &  persistence length \\
      $\theta$ & $\lp$ changes by a factor $1/\theta>0$ at $t=0$.  \\
      $k_B T$ & thermal energy \\
      $f(s,t)$ & line tension \\
      $\lpe(t)$ &  equilibration scale for transverse bending modes
      \\ 
      $\lpa(t)$ & scale  of tension variations \\
      $\vec \fe$ & external force \\
      $\vec  \gex$ & external force per arc length \\
      $\vec \xi(s,t)$ & thermal force per arc length
    \end{tabular}
  \end{ruledtabular}
\end{table}

\section{Elastic forces}
\label{sec:elastic-forces}

From the effective free energy of a worm-like chain, Eq.~(\ref{eq:bending-energy}),
we seek to determine the elastic force $\gel(s)$ per arc length that a polymer of a
given conformation $\bar{\vec r}(s)$ exerts at arc length $s$ onto its surroundings.
To this end, let us assume the polymer was subject to a constant external force field
equal to $-\gel(s)$, the negative of the local elastic forces.  Then, the
conformation $\bar{\vec r}(s)$ is in balance with the external force and minimizes
the total free energy
\begin{equation}
  \label{eq:Htot}
  \Htot[\vec r]=\Hb[\vec r]+\Hext[\vec r] 
\end{equation}
with
\begin{equation}
  \label{eq:Hext}
  \Hext[\vec r]\equiv \mint{ds}{0}{L}\gel(s) \,\vec r(s) \;,
\end{equation}
under all possible paths that obey the local inextensibility,
Eq.~(\ref{eq:inextensibility}). Hence, requiring a vanishing free energy change
$\delta \Htot$ for an infinitesimal (permitted) change $\delta \vec r$ in the space
curve leads to the sought-after relation between the elastic forces and the contour.
The central question of the minimization problem is how to deal with the local
constraint.

Here, we show that the common~\cite{goldstein-langer:95} introduction
of a Lagrange multiplier function ensuring the local inextensibility
constraint is not necessary.  We will instead present a minimization
procedure that considers only variations that obey the inextensibility
constraint to leading order.

Consider a test contour
\begin{equation}
  \label{eq:tcontour}
  \vec r(s)=\bar{ \vec{r}}(s)+\delta\vec  r(s)\;,
\end{equation}
which is infinitesimally displaced by $\delta \vec r(s)$ from the equilibrium contour
$\bar {\vec{r}}(s)$. The inextensibility constraint, Eq.~(\ref{eq:inextensibility}), is
fulfilled to $O(\delta \vec r)$ if we only consider displacements that are
constructed from another infinitesimal vector field $\vec \epsilon(s)$ according to
\begin{equation}
  \label{eq:transvers-variation}
  \delta\vec r'(s)=\vec \epsilon(s) \times \bar{\vec r}'(s)\;.
\end{equation}
Those displacements are transverse to the local tangent vector, so
that $\delta(\vec r'^2)$ is quadratic in $\delta \vec r$.  They
correspond to local rotations of the tangents.

The variation of the contour induces a variation $\delta {\cal H}$
of the total elastic energy ${\cal H}=\Hb+\Hext$ of the form
\begin{subequations}
  \begin{eqnarray}
    \delta {\cal H}&=& \mint{ds}{0}{L} \left(\kappa\, \bar {\vec{r}}''
      \delta\vec r''+\gel \delta \vec r \right)
    \label{eq:venergy}\\
    &=& -\mint{ds}{0}{L} \left(
      \kappa\, \bar {\vec{r}}'''+\mint{d\tilde{s}}{0}{s} \gel \right) \delta \vec r'+\text{b.t.}
    \label{eq:venergy-2}\\ 
    &=&-    \mint{ds}{0}{L} \left( \kappa\, \bar{ \vec{r}}'''+\fel \right) \left(\vec
      \epsilon
      \times \bar{\vec{r}}'\right)+\text{b.t.} \label{eq:venergy-3} \\
    &=&- \mint{ds}{0}{L} \vec \epsilon \left[ \bar{ \vec{r}}' \times \left( \kappa\,
        \bar{\vec{r}}'''+ \fel \right) \right]+\text{b.t.} 
    \label{eq:venergy-4}
  \end{eqnarray}
\end{subequations}
Here, we performed a partial integration to obtain Eq.~(\ref{eq:venergy-2}), which
introduces some boundary terms abbreviated by ``b.t.''. In the subsequent step we
inserted Eq.~(\ref{eq:transvers-variation}) for $\delta \vec r'$ and introduced the
arc length dependent force
\begin{equation}
  \label{eq:g-integrated}
  \fel(s)\equiv \mint{d\tilde{s}}{0}{s} \gel(\tilde{s}) \;.
\end{equation}
Finally, we used the property $\vec A\cdot(\vec B\times\vec C)=\vec
C\cdot(\vec A\times \vec B)$ of the triple scalar product to obtain
Eq.~(\ref{eq:venergy-4}).

If $\bar{ \vec{r}}(s)$ is indeed the equilibrium contour then $\delta {\cal H}$ has
to vanish for all variations parameterized by $\vec \epsilon(s)$ and $\delta \vec
r(L)$. Therefore, the term in the square brackets of the integrand in
Eq.~(\ref{eq:venergy-4}) has to vanish,
\begin{equation}
  \label{eq:beq-app}
  \bar{ \vec{r}}' \times \left( \kappa\, \bar {\vec{r}}'''+\fel
      \right)= 0 \;,
\end{equation}
so that we recover Eq.~(\ref{eq:beq}), used in Sec.~\ref{sec:dwmc}. In addition, the
boundary terms
\begin{subequations}
  \begin{eqnarray}
    \text{b.t.}&=& \kappa\left(\bar{ \vec{r}}'' \delta \vec r'\right)\vert^L_0+
    \delta \vec r(L)\mint{d\tilde{s}}{0}{L} \gel \nonumber\\
    &=&\kappa \left[\bar {\vec{r}}'' \left(\vec \epsilon \times
        \bar{\vec{r}}'\right)\right]\vert^L_0 +\left[\fel\delta\vec r\right]\vert^L
    \nonumber \;.
    \label{eq:venergy-5}
  \end{eqnarray}
\end{subequations}
have to cancel, implying the requirements
\begin{eqnarray}
  \fel(L)&=&0 \label{eq:tot-force} \\
  \left( \bar {\vec{r}}' \times \bar{ \vec{r}}''\right)\vert_{0,L}&=&0 \;.
   \label{eq:bc-free-:(}  
\end{eqnarray}
The condition expressed by Eq.~(\ref{eq:tot-force}) simply states that a force
balance can only exist if the external forces sum up to zero. Using the
inextensibility constraint, Eq.~(\ref{eq:inextensibility}), it is seen that $\bar
{\vec{r}}''\cdot \bar{ \vec{r}}'=0$, so that Eq.~(\ref{eq:bc-free-:(}) can be
rewritten as the boundary condition
\begin{equation}
  \label{eq:bc-free}
   \bar{ \vec{r}}''\vert_{0,L}=0\;.
\end{equation}
The curvature, which is proportional to the local torque, has to vanish at the (free)
ends.

Note that the over-bar in $\bar{\vec{r}}(s)$ to denote the equilibrium contour is
dropped in the main text, for simplicity.

\section{Multiple scale analysis (details)}
\label{sec:app-mspt}

\subsection{Thermal forces}
\label{sec:noise-mspt}

In the multiple scale perturbation theory, presented in Sec.~\ref{sec:multi-scale},
one should, in principle, assume an expansion $\vec \xi(x,y;t)=\epsilon^{1/2}\vec
\xi_1(x,y,t)+o(\epsilon)$ where the leading order noise $\vec \xi_1$ is a function of
the coarse-grained variable $y$.  However, from the fundamental correlations obeyed
by the original noise function, Eq.~(\ref{eq:noise-corrs}), we have to require to
leading order
\begin{equation}
  \label{eq:mspt-noise-bc}
  \begin{split}
    &\avg{\vec \xi_{1}(s,\epsilon s;t) \vec \xi_{1}(s',\epsilon s';t)}=2 (\bold I/L)
     \delta(s-s')\delta(t-t')
  \end{split}
\end{equation}
for \emph{all} $\epsilon\equiv L/\lp$. Apparently, the right hand side of
Eq.~(\ref{eq:mspt-noise-bc}) does not depend on $\epsilon$. As a consequence, the
left hand side cannot depend on $\epsilon$ either, in particular not on $\epsilon s'$
or $\epsilon s$. Hence, the two-point correlations are independent of the slowly
varying arc length coordinate.  Using Wick's theorem, we may argue in the same way
for all higher-order correlation functions as well and conclude that the leading
order stochastic force $\vec \xi_1$ is itself independent of $y$.

We note, that the inverse length appearing on the right hand side of
Eq.~(\ref{eq:mspt-noise-bc}) is due to the definition $\epsilon\equiv L/\lp$ of the
small parameter for a stiff polymer. In the case of a strongly pre-stretched
polymer, the small parameter is defined as $\epsilon\equiv (\lp\fe)^{-1/2}$, see
Eq.~(\ref{eq:wbl-sfp-2}) and the subsequent paragraph, so that $L^{-1}$ in
Eq.~(\ref{eq:mspt-noise-bc}) has to be replaced by $\fe^{-1/2}$.

\subsection{Next to leading order tension}
\label{sec:f1}

Requiring the $\Ord{\epsilon^{1/2}}$ coefficient in
Eq.~(\ref{eq:multiscale-2}) yields a noisy first order tension
\begin{equation}
  \label{eq:fi-1}
  f_1(x,y)=\mint{d\tilde
    x}{0}{x}\left[\xi_{\pa,1}(0)-\xi_{\pa,1}(\tilde x)\right] +b_1(y) x+a_1(y)  \,.
\end{equation}
For $f_1$ to be bounded in $x$, the term $b_1(y)x$ has to cancel the
linearly growing the noise-term on the right hand side,
\begin{equation}
  \label{eq:fix-c-1}
  b_1=\lim_{x\to\infty}\frac1x \mint{d\tilde
    x}{0}{x}\left[\xi_{\pa,1}(0)-\xi_{\pa,1}(\tilde x)\right]\;.
\end{equation}
However, important for the multiple scale analysis in
Sec.~\ref{sec:multi-scale} is merely that $b_1$ is independent of $y$,
so that $\partial_y\partial_x f_1(x,y)=0$.

\bibliographystyle{apsrev}

\bibliography{bibis/tdpre06,bibis/elastic-rod-dynamics,bibis/journals,bibis/sfpdynamics-tcited,bibis/unpub,bibis/actin-viscoelastic,bibis/semiflexibleA04,bibis/sf,bibis/klaussf,bibis/mysf,bibis/sfnet,bibis/unpub,bibis/mysfnet}

\end{document}